\documentclass[12pt]{article}
\usepackage{amssymb, amscd, amsmath, amsthm, epsf, psfig,latexsym}
\usepackage[dvips]{graphicx}

\usepackage{amsfonts}

\newtheorem{thm}{Theorem}[section]

\newtheorem{lemm}[thm]{Lemma}

\newtheorem{p}[thm]{Proof}

\newtheorem{CO}{Corollary}

\newtheorem{OB}{Observation}

\newcommand{\C}{{\mathbb C}}

\newcommand{\RI}{{\rightharpoonup}}

\def\U{{\bf U}}

\begin{document}

\thispagestyle{plain} \rule{0cm}{25.2mm}\\
{\noindent{\Large \bf Poly-locality in quantum computing}}

\vspace{5.2mm}
 Michael H. Freedman\footnote{\small Electronic mail: michaelf@microsoft.com}

\vspace{1mm}

{\small{\it Microsoft Research, One Microsoft Way, Redmond, WA
98052 U.S.A}}

\vspace{5.5mm}

\begin{abstract}
A polynomial depth quantum circuit effects, by definition a
poly-local unitary transformation of tensor product state space.
It is a reasonable belief [Fy][L][FKW] that, at a fine scale,
these are precisely the transformations which will be available
from physics to solve computational problems.  The poly-locality
of discrete Fourier transform on cyclic groups is at the heart of
Shor's factoring algorithm.  We describe a class of poly-local
transformations, which include the discrete orthogonal wavelet
transforms, in the hope that these may be helpful in constructing
new quantum algorithms.  We also observe that even a rather mild
violation of poly-locality leads to a model without one-way
functions, giving further evidence that poly-locality is an
essential concept.
\end{abstract}

\section{Introduction}

Feynman [Fy] described how, in principle, a finite dimensional
quantum mechanical system could be made to evolve so as to execute
a sequence of $\lq\lq$quantum gates."  By a quantum gate, we mean
one of the several rather standard unitary operators applied to a
small (usually 1, 2, or 3) number of tensor factors (and the
identity on the remaining factors) within a larger tensor product
of factors of bounded dimension, often taken to be
$(\C^{2})^{\otimes n}$.  For example: NOT, Controlled NOT, and
Controlled $i-$phase is known to be a complete basis [Ki].

Lloyd [L] made the converse  explicit. Let $H$ be a poly-local
Hamiltonian, $H =\overset{L}{\underset{\ell = 1}{\sum}} H_\ell$ is
a sum of $ L \leq$ polynomial(problem size) terms $H_\ell$, each
the identity except for a constant number of tensor factors.  The
Hamiltonian $H$ exponentiates to $\U = e^{2 \pi i t H}$ which is
well approximated by $\U^{\prime}=\overset{M}{\underset{i =
1}{\prod}} (\text{gate})_i$, where $M \leq \text{constant}\,\, t
\cdot \frac{1}{\epsilon} \cdot L^2$ and $\epsilon > 0$ is the
allowed (operator norm) discrepancy between $\U$ and $\U'$.  The
poly-local assumption on $H$ is very natural for most real world
quantum mechanical systems and is often summarized by the axiom:
$\lq\lq$physics is local."  For example, in quantum field theory
interactions tensors with more than four indices are generally not
renormalizable.  In more exotic contexts, such as, fractional
quantum Hall systems [W] topological models have been proposed for
which the Hamiltonian acting on internal degrees of freedom
vanishes identically $H \equiv 0$, yet after a topological event,
such as a braiding $b$ is completed, the internal state is
transformed by $\U(b)$.  For systems of this type (called: a
unitary topological modular functor) a distinct argument was
provided [FKW] that $\U(b)$ is poly-local.  There seem to be few
remaining candidates - perhaps quantum gravity - for a physical
system having a finite dimensional state space whose evolution
cannot be well approximated by a poly-local transformation $\U$.
This reinforces confidence in the quantum circuit model QCM [Ki],
and motivates the examination of (families of) transformations
having a poly-local description. Any poly-local unitary transform
is available as a $\lq\lq$subroutine" on a quantum computer, but
it will not generally be a simple matter to recognize a
transformation as lying in (or efficiently approximated by) that
class.

While a simple volume argument shows that only a tiny fraction of
the unitary group $\U(2^{n})$ can be approximated (up to phase) by
poly-local transformations, it is not known how to give concrete
examples which cannot be thus approximated.  The situation is
reminiscent of the fact that transcendental numbers are easily
seen to predominate, but less easy to identify.  To
add to the puzzle, in most contexts, we will be free to add
ancilli to create an inclusion $v \mapsto v \, \otimes \mid 0
\dots 0 >$, $H \subset H^{+}$ of one Hilbert space into a
larger one.  It seems likely that there may be an operator $T:H
\longrightarrow H$ which is not well approximated by poly-local
operators but its extension $T \otimes \text{id} : H^{+}
\longrightarrow H^{+}$ is.  In this regard, Michael Nielsen has
pointed out (private communication and Chpt. 6 [CN]) that the
exponential of any Hamiltonian which is a sum of polynomially many
product terms can, with the help of ancilli, be efficiently
simulated on a quantum computer.  There is no requirement on the
individual product terms except that they be products, i.e. a
tensors of operators each acting on one (or up to log many)
qubit(s).

To appreciate the importance of a poly-local description consider the
FFT.  The reader familiar with the Fast Fourier Transform may be
surprised that it has a virtue in the quantum context since it is
already fast, $\mathcal{O}(N \log N)$, classically.  But on this
scale, we should think of the quantum version as poly$(\log N)$ in
the sense that we may Fourier transform in poly$(\log N)$ time a
data vector $\bigl( f(\mid 1 >), \dots, f(\mid {N}>)\bigr)$ (held
in superposition and previously computed in poly$(\log N)$ time)
into its discrete Fourier transformation $\bigl( \widehat{f}(\mid
\widehat{1} >), \dots, \widehat{f}(\mid \widehat{N} >)\bigr)$. The
measurement phase of quantum computation samples the transform
according to the density of its $L^2 -$norm squared.  In a sense,
this is the best we could hope for in poly-log time.  We could
never output the entire transform in less than $\mathcal{O}(N)$
time, instead sampling shows us its sharpest spikes. Similarly
with other transforms, a poly-local description allows us to
sample, according to the $L^2 -$ norm, that transform applied to
any rapidly computable function.

A key step in the Shor factoring algorithm is his description of
the discrete Fourier transform $F: C_m \longrightarrow
\widehat{C}_m$ for the cyclic group of order $m$, where $m$ is a
large smooth number, as a poly-local transform (The size parameter
is $\log m$ so $F$ must be written as a composition of poly$(\log
m)$ many gates).

We construct another class of examples by showing that cyclic
periodic near diagonal matrices are poly-local (Thm 1).  The
construction is direct - no ancilla qubits are used. As an
application, we prove (Corollary 1) that all the standard
orthogonal discrete wavelet transforms, in particular the
Daubechies transforms $D_{2n}$ [D], are poly-local.  I thank C.
Williams for pointing out that he and A. Fijany previously
obtained an explicit factoring of $D_2$( = Haar) and $D_4$ into
quantum circuits [AW].  Our alternative approach has the usual
virtues and demerits of abstraction.

Some perspective into the limitations poly-locality imposes on the
poly-time computational class of the QCM, BQP, can be achieved by
adjoining \underline{oracles}.  Let $\mathcal{O}_{n}:
(\C^{2})^{\otimes n} \longrightarrow  (\C^{2})^{\otimes n}$ be a
$\lq\lq$super-gate" which can be called upon in addition to gates
to transform the computational state $\Psi$.  It is clear that if
$\mathcal{O}_n$ encodes the answer to a difficult (or even
undecidable) problem $P_n$, $\text{QCM}^{\mathcal{O}}$ will be
improbable strong:  If $f_n : \{0,1\}^{n-1} \longrightarrow
\{0,1\}$ is a function and $\wedge (f_n):\{0,1\}^{n}
\longrightarrow \{0,1\}^{n}$ is defined by $\wedge f_n (x_1 ,
\dots, x_{n-1},x_n) = \{x_1 , \dots, x_{n-1}, f_n (x_1 , \dots,
x_{n-1}) + x_n \}$ and $\mathcal{O}_n$ is the linear extension of
$\wedge f_n$ operating on $(\C^2)^{\otimes n}$, then  we can
compute $f_n (x_1 , \dots, x_{n-1})$ by measuring the last qubit
of $\mathcal{O}_n (x_1 , \dots, x_{n-1}, 0)$.   However, if we
stipulate that $\mathcal{O}_n$ is $\lq\lq$locally-poly" meaning
that the $(i, j)$ entry of $\mathcal{O}_n$ be computable in
poly$(n)$ time, can the nonlocal structure of such $\mathcal{O}_n$
by virtue of their size alone, extend QCM?  The following
observation is evidence that the answer is $\lq\lq$yes."

\begin{OB} For each family of bijections $f_n:\{0,1\}^{n} \longrightarrow
\{0,1\}^{n}$ which is computable in polynomial(n) time there is a
family of oracles $\mathcal{O}_n$ with entries $(\mathcal{O}_n
)_{i, j}$ computable in poly$(n)-$time so that $f^{-1}$ can be
computed by $\text{BQP}^{\mathcal{O}_n}$.  Simply set
$(\mathcal{O}_n )_{i, j} = \delta_{f{(i), j}}$. Write the bit
string $(y_{1}^{j}, \dots ,y_{n}^{j}) = \overset{\RI}{y} = y^{j},
1 \leq j \leq 2^{n}$. Then $\mathcal{O}_{n} \overset{\RI}{y} =
\overset{2^{n}}{\underset{j=1}{\Sigma}} \delta_{f(i), j}\, y^{i} =
y^{f^{-1}(j)} = f^{-1} (\overset{\RI}{y})$, so applying
$\mathcal{O}_n$ to $(y_{1}, \dots , y_{n})$ and then measuring
each qubit separately solves the inversion problem
$\mathcal{O}_{n} (y_{1}^{j}, \dots , y_{n}^{j}) = \big(
y_{1}^{f^{-1} (j)}, \dots , y_{n}^{f^{-1} (j)} \big)$. Thus,
one-way functions disappear if we are permitted to adjoin
locally-poly yet non-poly-local oracles. We have not succeeded in
solving an $NP$ complete problem with such an oracle.
\end{OB}

\section{Periodic Band-Diagonal Transforms}

It is common in signal processing to encounter smoothing operators
which when written as matrices are band diagonal.  Often
mathematical reasons suggest wrapping the interval into a circle
and considering cyclically banded matrices $M$, $M_{i, j}=0$
unless $\mid i-j \mid < b \mod N, 1 \leq i, j \leq N$.  Two
additional conditions natural in signal processing are unitary,
$(M^{-1}=M^{\dagger})$ and cyclicity $\bigl( \exists k << N \,\,
\text{s.t.}\,\, M_{i, j}= M_{i+k,\,\, j+k}$, addition of indices
taken $(\text{mod} N) \bigr)$. All these conditions occur in
\underline{quadrature mirror filters} [P].  We fix $b$ and $k$ and
considering $N$ as parameter approaching infinity for studying the
family of $N \times N$ matrices $M_{N}$.  To summarize the $M_{N}$
are $N \times N$ unitary matrices, $b-$band diagonal {\it in the
cyclic sense}, cyclic with period $k$, $k|N$, and $4b<N$.
Furthermore even among different $M_{N}$, $M_{N^{\prime}}$, we
assume $(M_{N})_{i,j}=(M_{N^{\prime}})_{i^{\prime},j^{\prime}}$.
If $i \equiv i^{\prime}$ and $j \equiv j^{\prime} \mod k$, so the
matrices look locally identical.  Purely for convenience, we also
assume all $N$ are powers of $2$, $N= 2^{n}$, $n > n_\circ$ some
fixed constant and $k=2^\ell$, this makes the fit with qubits
effortless. A final technical assumption is necessary to
efficiently approximate even a small $M_{N_\circ}$, $N_\circ =
2^{n_\circ}$, from gates available in the QCM.  It is sufficient
here [Ki] to require the entries $M_{i, j}$ are algebraic numbers.
This prevents some difficult to obtain information being encoded
in the entries themselves.  The theorem below asserts that the
family $\{M_{N}\}$ is poly-local in the strictest sense:  No
ancilla are introduced to stabilize the problem.

\begin{thm}
 Let $\{M_{N}\}$ be as above, there is a polynomial $p$ so
that for any $\epsilon > 0$ there is an $M'_{N}$, $\epsilon$ close
to $M_{N}$ in the operator norm, so that $M'_{N}$ is a composition
of $p \Bigl( n, \log \bigl( \frac{1}{\epsilon} \bigr) \Bigr)$
gates in the QCM.
\end{thm}

\begin{p}
\textnormal{It is sufficient to consider $N = 2^{n}$, for $n$
greater than any fixed number, so the reader may image $M_{N}$,
$N$ comfortably large.  For some $K = 2^{j} k$, some $j>0$, we
will modify $M_N$ to $\overline{M}_{N}$ to a unitary $K -$cyclic
matrix which agrees with $M_N$ except $K-$periodically for a small
(order $b$) slab of rows where $\overline{M}_{N}$ agrees with the
identity matrix and a small (order $b$) transition region between
the two types of rows.}
\end{p}

\begin{figure}
  \centering
\includegraphics[width=11cm, height=6cm]{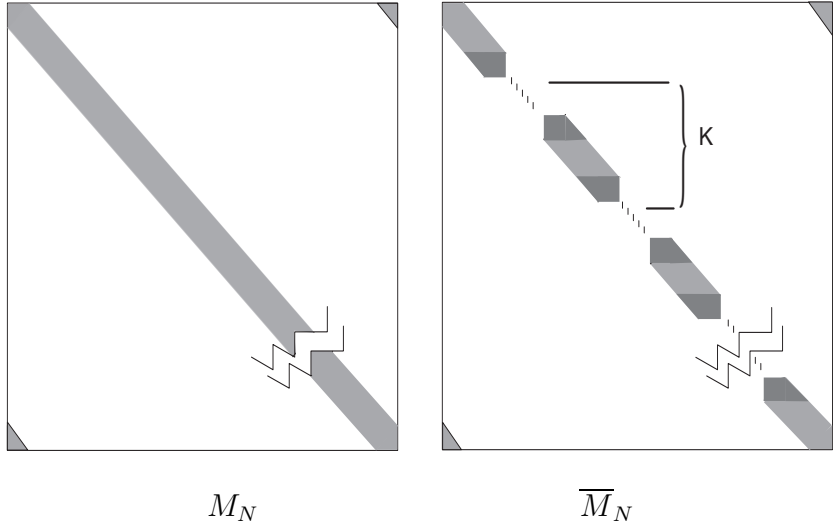}
 \center{$M_N$ \qquad \qquad \qquad \qquad \qquad $\overline{M}_N$}
  \caption{ White = zero, light shading = original K-periodic entries,
  dark shading = transition rows, and 1 = one. }
\end{figure}

The construction of $\overline{M}_{N}$ is very similar to the
truncation process given in [F,P] so we will not give formulas
here, but simple principles from which $\overline{M}_{N}$ may be
constructed.  Let $V_I$ be the vector subspace spaned by the rows
of $M_{N}$ whose row index lies in $I$ (We think of $I$ as an
interval with integer end points on the circle consisting of
$R/NZ$).  Let $W_I$ be the subspace spanned by the basis vectors
with indexes in $I-$equivalently those rows of the identity $N
\times N -$matrix.  From the nonsingularity of unitary matrices we
obtain:

\begin{lemm}
Whenever $J$ contains the $b-$neighborhood of $I$, $W_{I} \subset
V_J$ and $V_{I} \subset W_J. \,\, \square$
\end{lemm}

We apply the lemma several times below:  consider an interval $I$
of $\ell >2b$ rows and a larger interval $J$ of $\ell \,\,
\text{rows} +2b$ rows containing $I$ symmetrically. Let $J^{+}$
and $J^{-}$ denote the intervals of length$= 2b$ at containing the
$+$ and $-$ end points of $J$ respectively. The rows $\{q\}$ of
$M_N$ in the complement of $I$ together with the rows $\{s\}$ of
the identity matrix in the interval $J$ span the entire space. Let
$\{r\}$ denote the rows of $M_N$ with indices in $J^{+}$ and
$J^{-}$.  The Gram-Schmidt process applied to $\{r\} \cup \{s\}$
(The Gram-Schmidt process requires an ordering; work from the
middle of the $J$ interval outward) produces a Hermitian-
orthogonal frame $\{t\}$ for the space spanned by $\{r\} \cup
\{s\}$ (Note: $\{q\} \cup \{r\} \cup \{s\}$ span the entire
space).  This process naturally leaves the central $\ell-2b$ rows
of $\{s\}$ unchanged, i.e. they remain standard basis vectors. The
geometry of this process is pictured below.

\vskip.1in \epsfxsize=3.5in \centerline{\epsfbox{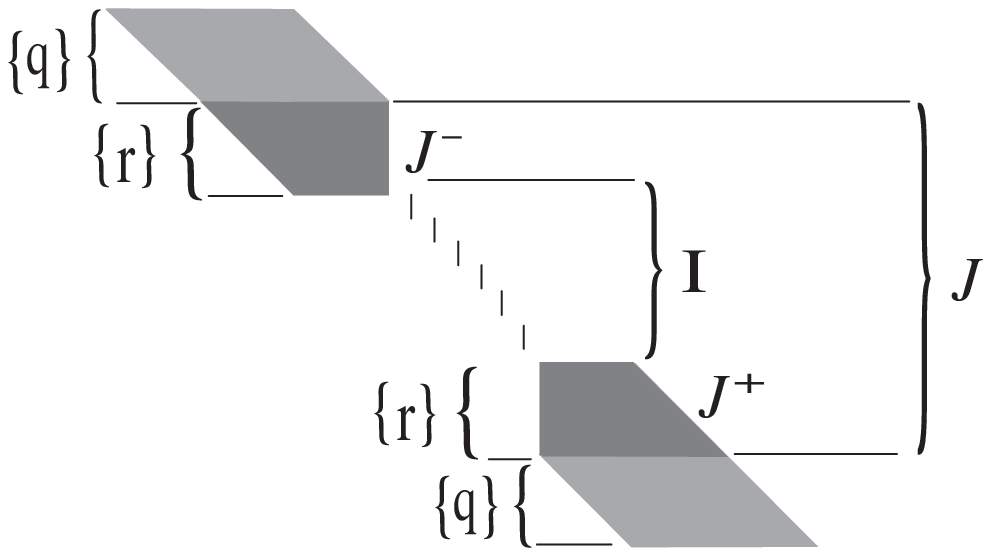}}
\centerline{Figure 2} \vskip.1in

The frame $\{t\}$ consists of $\ell$ standard basis vectors
together with $2b$ vectors of length $\leq 2b$.  Think of these
$2b-$vector as mediating a transition from rows $\{q\}$ of $M_N$
to rows  $\{s\}$ of the identity matrix.  The truncation process
preserves unitarity and band diagonality and the band width.

$\overline{M}_N$ is block diagonal consisting of $2^{n-j-\ell}$
identical blocks $\overline{M}$ (independent of N) down the
diagonal.  Thus, $\overline{M}_N$ has a tensor decomposition
$\overline{M}_N = \overline{M} \otimes \text{id}_{(\C^2)^{\otimes
n-j-\ell}}$.

Define $\overline{\overline{M}}=(\overline{M}_{N})^{-1} M_N$.
Clearly the rows of $\overline{\overline{M}}_N$ agree with the
identity matrix outside the $J-$intervals and
$\overline{\overline{M}}_N$ is $b-$band diagonal.  Thus,
$\overline{\overline{M}}_N$ can be written in a tensor product
form as well: $\overline{\overline{M}}_N$ consists of identical
blocks down the diagonal of size $K$ provided $K \geq 2b$. Thus,
we may write $\overline{\overline{M}}_N=P \circ(
\overline{\overline{M}} \otimes \text{id}_{(\C^2)^{\otimes
n-j-\ell}}) \circ P^{-1}$ where $P$ is a cyclic permutation matrix
of $\{1, 2, \dots, N\}$.  The permutation must arise since the
blocks of $\overline{\overline{M}}$ are off-set from the blocks
$\overline{M}_N$ by a translation of $\frac{K}{2}$.

Because we assumed the entries of $M_N$ to be algebraic so are the
entries of $\overline{M}$ and $\overline{\overline{M}}$.  It is
well-known that algebraic numbers can be efficiently computed
using Newton's method so that refining the numerical accuracy of
an algebraic $\alpha$ to up to a factor of $(1 + \epsilon)$
requires only constant$(\alpha) \cdot$ poly
$\log\bigl(\frac{1}{\epsilon}\bigr)$ computational steps.  Now by
a theorem of [Ki] for approximating a unitary transformation in a
fixed dimension as a product of gates, we find matrices
$\overline{M}^{\prime}$ and $\overline{\overline{M}}^{\prime}$, $|
\overline{M} - \overline{M}^{\prime}|< \frac{\epsilon}{2}$, and $|
\overline{\overline{M}} - \overline{\overline{M}}^{\prime}|<
\frac{\epsilon}{2}$ using the operator norms, with
$\overline{M}^{\prime}$ and $\overline{\overline{M}}^{\prime}$ a
product of poly $\log \bigl( \frac{1}{\epsilon} \bigr)$ elementary
gates.

The cyclic permutation $P$ and its inverse $P^{-1}$ are, of
course, poly$(n)=\text{poly}(\log N)$ time computable classically.
Writing $|x > \in \{1,2,\dots, N \}$ in an initial register and
$|0>$ in an equal large (size $=n$, $N = 2^{n}$) register of
ancilli a well-known sequence of operations effects the
transformation:
\begin{gather*}
\begin{matrix}
P_{i} (| x> \otimes |0>) \,\, \overrightarrow{\text{compute} P}
\,\, \bigl( |x> \otimes |\,0 + P(x) > \bigr) \,\,
\overrightarrow{\text{flip}} \,\,
 \bigl(| P(x)> \otimes | x > \bigr) \\
\overrightarrow{\text{compute} P^{-1}} \,\,
 \bigl( | P(x) > \otimes
| x + P^{-1} P(x) > \bigr) \\ = \big( |P(x) > \otimes |0 > \big)
\end{matrix}
\end{gather*}
(where $+$ denotes bit wise mod $2$ sum in the register). More
directly, it is possible to build size $\mathcal{O}(n^2)$ circuits from
$\lq\lq$NOT, $\lq\lq$Controlled NOT," and $\lq\lq$swap" gates (See Section 2.4 [AW].), which use no
ancilli and implement addition of $\frac{K}{2}$ modulo $N$.  For
example, a circuit implementing $+2 \mod (8)$ is shown below.

\vskip.2in \epsfxsize=4in \centerline{\epsfbox{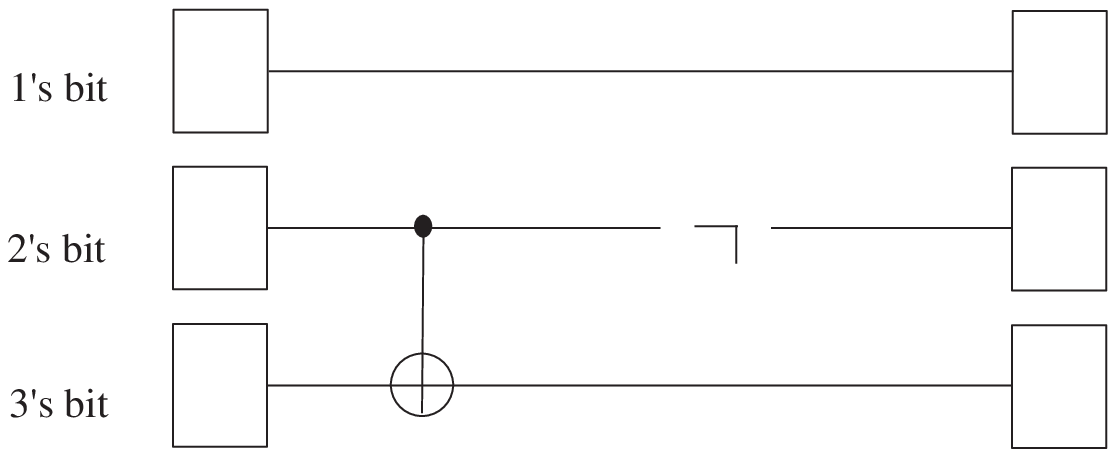}}
\centerline{Figure 3} \vskip.2in

\noindent Thus, all the 4 terms of an $\epsilon-$approximation
$M^{\prime}_N$ may be written as a product of poly $(n, \log
\frac{1}{\epsilon})$ many gates: $M_N \overset{\epsilon}{\approx}
M^{\prime}_N  = (\overline{M}^{\prime}\otimes \text{id}) \circ P
\circ (\overline{\overline{M}^{\prime}} \otimes \text{id}) \circ
P^{-1}$. $\square$

\section{Wavelets}

    We define a discrete wavelet transform as the result of the
\underline{pyramid algorithm} [M] - applied to a band diagonal
quadrature mirror filter.  Daubachies' [D] original basis is
recovered by choosing a quadrature mirror filter whose \\
\underline{differencing} \underline{rows} have a maximal number of
vanishing moments (consistent with orthogonality relations). For
example, the simplest Daubachies wavelet Daub$_{4}$ is built from
an orthogonal matrix(and mirror filter):
\begin{gather*}
M_N= \overbrace{
 \begin{vmatrix}
    C_1 & C_2 & C_3 & & & & &C_0 \\
     -C_2 & C_1 & -C_0 & & & & & C_3\\
    &C_0 & C_1 & C_2 & C_3 \\
    &C_3 & -C_2 & C_1 & -C_0\\
    \vdots & & \vdots &&&& \vdots\\
    &&&C_0 & C_1 & C_2 & C_3\\
    &&&-C_3 & C_2 & C_1 & -C_0\\
    C_3&&&&&C_0 & C_1 & C_2\\
    -C_0&&&&&-C_3 & C_2 & C_1\\
  \end{vmatrix}}^{2^n = N} \Biggr\} 2^n = N
\end{gather*}
satisfying moment conditions:
\[C_3 - C_2 + C_1 - C_0 =0 \;\; \text{and}\;\;
0C_3 - 1C_2 + 2C_1 - 3C_0 = 0.\]

First let us explain in algorithmic terms how the
\underline{wavelet coefficients} of an data (column) vector
$X=(x_{1}, \dots, x_{N})$ are computed starting from a filter
$M_N$ as above (compare [P]). The even components of $M_N X$ are
regarded as $\lq\lq$smoothed" and these are set aside for further
processing.  The odd components of $M_N X$ are the finest scale
wavelet coefficients.  To continue processing apply the matrix
identical in description to $M_N$, but of half the size to the
even output of $M_N X: M_{N/2} \bigl( \text{even} (M_N X)\bigr)$.
Again set the even output aside: the  odd output are the wavelet
coefficients at the next-to-smallest scale.  Now apply $M_{N/4}
\Bigl(\text{even} \bigl( M_{N/2}(\text{even} M_N X)\bigr) \Bigr)$,
etc...  The process must terminate when $M$ becomes so small that
wrapping around of rows threatens unitarity. For Daub$_{4}$ the
last $M$ will be a $4 \times 4$ matrix.  The wavelet coefficients
of $X$, usually written as $X$ integrated against a function
$\psi$, are the output of the final differencing (proceeded,
possible, by several smoothings).  The $\lq\lq$mother" or
$\lq\lq$scaling" coefficients will be the output of successive
smoothings (i.e. even rows). These are traditionally written as
$X$ integrated against a  function $\phi$; there will be two of
these for Daub$_4$.

The pyramid algorithm consist of only $\log(N)$ many applications
of a filter matrix $M$, obeying the hypotheses of the theorem, and
some easily computed permutations needed to sort out the even from
odd rows between applications of $M$.  Thus from the theorem, we
have an evident corollary, which should be compared with the constructions of
[AW] and [Kl].

\begin{CO}  Consider a poly-time computable function of $f$ from the
integers $[1, \dots , N]$ into a fixed basis of a vector space $V$
of dimension $=d$.  The function $f$ is recorded by a vector
$v_{f} \epsilon V^N$. For any discrete (algebraic) wavelet
transform $T$ applied to $f$, there is a quantum circuit, without
ancilla, of length poly$\big( \log N, \, \log \frac{1}{\epsilon}
\big)$ which yields a state space vector
$T^{\prime}(v_{f})\epsilon V^N$ which is within $\epsilon >0$ of
the exact transform $T(v_{f})$. The components of
$T^{\prime}(V_{f})$ can be sampled by von Neumann measurement
according to the density $|\text{component}_{i, j}|^{2}$, $1 \leq
i \leq N$, $1 \leq j \leq d$.
\end{CO}

The proceeding recipe, in so far as repeated smoothing is applied,
sends a $\lq\lq$Dirac function" $\delta$ toward a fixed point of
the iteration $\phi (x) \longrightarrow \Sigma c_n \phi (2 x-n)$.
Such a fixed function $\phi$ is called the scaling function of the
filter $M$. Wavelet coefficients (in the limit) correspond to
integrations against $\psi(2^{k} x-n)$ where
$\psi(x)=\Sigma(-1)^{k} c_{1-k} \phi (2x-k)$.  Whereas, the
Fourier transform efficiently extracts information about
periodicity (hence its relevance to divisibility and factoring
questions) the wavelet transforms (discrete or continuous) extract
qualitatively different information.  It is less easy to say
precisely what combinatorial structures wavelets are best suited
to detecting, but the answer must relate to phenomenon with a
certain amount of both spacial and frequency localization.  Some
studies suggest fractal behaviors, Hausdorff dimension, and the
like are readily observed in wavelet bases.  Thinking
democratically, one should not prefer either Fourier or wavelet
bases, but seek different problems which can be solved by sampling
the density of one transform or the other.  So far on the Fourier
side there is discrete log, factoring, and the abelian stabilizer
problem [S][Ki] - essentially similar applications of that
transform, and no interesting quantum applications, yet known, for
the wavelet transforms.

\section{Continuum Versus Vectorized Transforms}

Both Shors' [S] use of the Fourier transform and our treatment of
wavelet transforms concerns the $\lq\lq$vectorized" transform: The
function values of $x^{a} \mod(n)$ in Shor's work and the values
of $f$ mod$(d)$ in our corollary 1 are treated as independent
basis vectors in an abstract vector space.  The transforms do not
treat similar function values similarly, but rather place each
possible outcome of $f(t)$ into a separate bin.  Thus, there is no
notion of {\it the continuum limit} of these transforms as
presently constituted and they should not be confused with their
continuous relatives in signal processing.  The property of a
function $f$ on the integers having period $=d$ is
\underline{invariant} under arbitrary permutation (or
vectorization) of the target.  This fact is crucial to the
efficacy of the Fourier transform in Shor's algorithm.  It is more
difficult to see what portion of the information captured by a
wavelet transform is invariant under target permutation or
vectorization.

Although it seems that a continuous - as opposed to vectorized -
Fourier transform is \underline{not} as useful for factoring,
their may be other quantum applications for the continuous
version.  In the case of wavelets, it is more likely that
continuous rather than vectorized transforms will be important. In
the following paragraph, we explain how to use phase coding of
states to build efficient continuous versions of any transform $T$
(such as Fourier or wavelet) for which there is an efficient
quantum circuit implementation of the vectorized version.

Let $F: [1, \dots , N]\longrightarrow [1, \dots, d]$ be a funtion
and denote $t \in [1, \dots, N]$ as the variable.  For $1 \leq x
\leq d$, we may define $f_{x}(t)=0$ if $f(t) \neq x$, and
$f_{x}(t)=1$ if $f(t)=x;\,\,
f(t)=\overset{d}{\underset{x=1}{\Sigma}} f_{x} (t)$.  A
$\lq\lq$vectorized" unitary transform obeys:
\[T
\Bigl( \overset{N}{\underset{t=1}{\Sigma}} | t > \otimes | f(t)
{>} \Bigr) = \overset{d}{\underset{x=1}{\Sigma}}\,\,
\overset{N}{\underset{\widehat{t}=1}{\Sigma}} \Bigl( T
(f_{x})\widehat{t} \Bigr) | \widehat{t} {>}\otimes |x>\] where $1
\leq \widehat{t} \leq N$ is the transform variable, and states are
written without numerical normalization factors. If we pick $D
> > d$ and let $w = e^{2 \pi \, i/D}$, the set of complex numbers $S=
\{w^{i} |1 \leq i \leq d \}$ is a short discretized arc of a
circle and is a rather good affine model for $\triangle = [1,
\dots, d]$ under the bijection $\delta
\overset{b}{\rightarrowtail}e^{2 \pi i \, \delta/D}$. After linear
rescaling the quasi-isometric distortion of b is $\mathcal{O}
\Bigl( \big(\frac{d}{D}\big)^{2} \Bigr)$ (A map $g:X
\longrightarrow Y$ between metric spaces is said to have
quasi-isometric distortion $\leq K$ for some $K \geq 0$ iff:
\[ \frac{1}{1+K}\,\, d \big( g (x_{1}), g (x_{2}) \big) \leq d
(x_{1}, x_{2}) \leq (1 +K) d \big( g (x_{1}) g(x_{2}) \big).\] We
will neglect this distortion and code the values of a bounded real
function $h:[1, \dots , N] \longrightarrow \big[ 0, \frac{2 \pi
d}{D}\big]$ into $\widetilde{h} : [1, \dots, N] \longrightarrow
\bigtriangleup$, $\widetilde{h} = b^{-1} \circ r \circ \text{exp}
(ih)$ where $r$ rounds to the nearest element of the form $e^{2
\pi i \, \delta/D}$.  Now a continuous version $\widetilde{T}$ of
$T$ applied to $\widetilde{h}$ is given by the formula:
\[\widetilde{T} \Bigl( \overset{N}{\underset{t=1}{\Sigma}} | t{>}\otimes
|\widetilde{h}(t)
>\Bigr) = \overset{d}{\underset{x=1}{\Sigma}}\,\,
\overset{N}{\underset{t=1}{\Sigma}} w^{x}
\Bigl(\widetilde{T}(f_{x}) \widehat{t}\Bigr) |\widehat{t}>.\]

Note that there is a simple modification of a quantum circuit
computing $T$ to one computing $\widetilde{T}$.  Instead of simply
recording the qubit $|x>$ by adding $x$ to a $D-$state register
$R$ holding zero, the trick is to initialize $R$ to the
superposition:  \[w^{0}|1> + w^{1} |2 > + \dots w^{D-1}|D>.\] Now
adding $x$ to the above state simply changes its phase by a factor
of $w^{x}$.  Thus, the transform $\widetilde{T}$, possessing a
natural continuum limit, can be sampled in the QCM with no more
difficulty than the discrete version $T$.  If quantum computers
are built, such continuous transforms would find application in
signal processing.  Beyond these obvious applications, it is
interesting to wonder what algorithmic potential lies in these
continuum transformations.

\end{document}